\documentclass[aps, prl, reprint, superscriptaddress, nofootinbib]{revtex4-1}
\usepackage[utf8]{inputenc}
\usepackage[english]{babel}
\usepackage{physics}
\usepackage{times}
\usepackage{graphicx,epsfig}
\graphicspath{{images/}}
\usepackage{color}
\usepackage[usenames,dvipsnames]{xcolor}
\usepackage{amsmath,bbm,amssymb, amsthm}
\usepackage{dsfont} 
\usepackage{stmaryrd}
\definecolor{linkcolor}{rgb}{0,0,0.6}		
\definecolor{bleu}{HTML}{1732a6}
\usepackage[colorlinks=true, pdfstartview=FitV, linkcolor= linkcolor, citecolor= linkcolor, urlcolor= linkcolor, hyperindex=true, hyperfigures=false]{hyperref}
\usepackage{numprint}

\addto\captionsenglish{}

\begin{document}

\title{A two-qubit engine fueled by entangling operations and local measurements}

\author{L\'ea Bresque}
\affiliation{Universit\'e Grenoble Alpes, CNRS, Grenoble INP, Institut N\'eel, 38000 Grenoble, France }

\author{Patrice A. Camati}
\affiliation{Universit\'e Grenoble Alpes, CNRS, Grenoble INP, Institut N\'eel, 38000 Grenoble, France }

\author{Spencer Rogers}
\affiliation{Department of Physics and Astronomy, University of Rochester, Rochester, NY 14627, USA}

\author{Kater Murch}
\affiliation{Department of Physics, Washington University, St. Louis, Missouri 63130}

\author{Andrew N. Jordan}
\affiliation{Department of Physics and Astronomy, University of Rochester, Rochester, NY 14627, USA}
\affiliation{Institute for Quantum Studies, Chapman University, Orange, CA, 92866, USA}

\author{Alexia Auff\`eves}
\affiliation{Universit\'e Grenoble Alpes, CNRS, Grenoble INP, Institut N\'eel, 38000 Grenoble, France }

\begin{abstract}
We introduce a two-qubit engine that is powered by entangling operations and projective local quantum measurements. Energy is extracted from the detuned qubits coherently exchanging a single excitation. This engine, which uses the information and back-action of the measurement, is generalized to an $N$-qubit chain. We show that by gradually increasing the energy splitting along the chain, the initial low energy of the first qubit can be up-converted deterministically to an arbitrarily high energy at the last qubit by successive neighbor swap operations and local measurements. Modeling the local measurement as the entanglement of a qubit with a meter, we identify the measurement fuel as the energetic cost to erase correlations between the qubits.
\end{abstract}

\maketitle

Understanding quantum measurements from a thermodynamic standpoint is one of the grand challenges of quantum thermodynamics, with strong fundamental and practical implications in various fields ranging from quantum foundations to quantum computing. Quantum measurement has a double status: on one hand, it is the process that allows the extraction of information from a quantum system. In the spirit of classical information thermodynamics, its ``work cost" was thus quantitatively analyzed as the energetic toll to create correlations between the system and a memory \cite{Parrondo2015,SagawaUeda,Jakobs}.  On the other as  stochastic processes, quantum measurements also lead to wavefunction collapse. Measurements can thus behave as a source of entropy and energy, playing a role similar to a bath.  The energetic fluctuations generated by the measurement backaction have recently been exploited as a new kind of fuel in so-called measurement-driven engines \cite{Elouard_maxwell, Elouard_Jordan,Yi2017,ding2018measurement,JEA,Mohammady2017}, and quantum fridges \cite{Campisi, Buffoni}.

\begin{figure}[ht]
	\centering
	\includegraphics[width=1\linewidth]{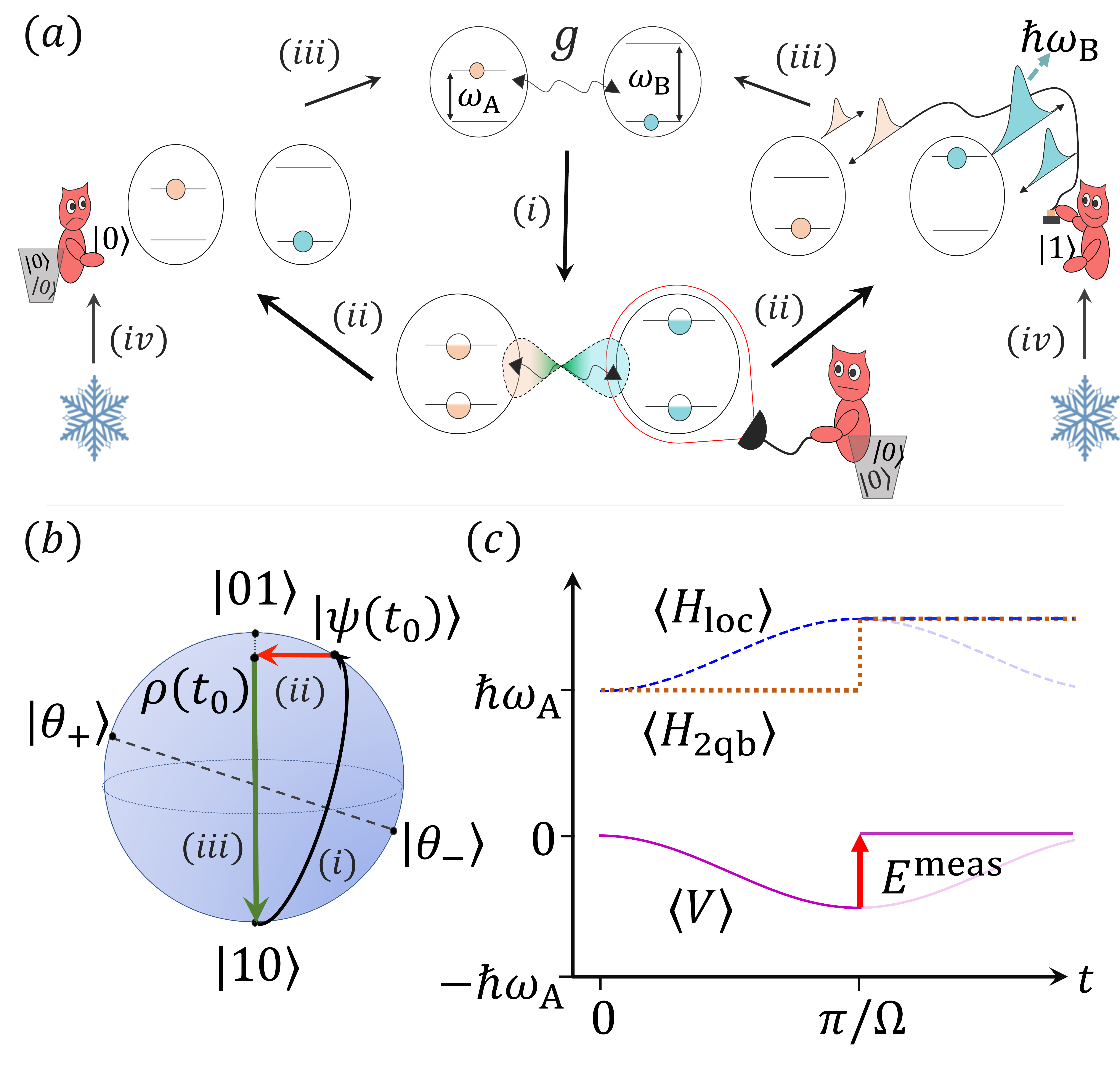}
	\caption{A two-qubit engine.  (a) Scheme of the engine cycle. \textit{(i)} Starting from $\ket{10}$, the qubits get entangled by coherently exchanging an excitation. \textit{(ii)} A demon performs an energy measurement on qubit $B$ at $t_0=\pi/\Omega$. \textit{(iii)} Feedback. If $B$ is found in the excited state, a $\pi$ pulse is applied to each qubit. The energy of $B$ is extracted and $A$ is re-excited. If not, nothing is done. At the end of this step, the qubits are back to their initial state. \textit{(iv)} Reset of the demon's memory. (b) Representation of the qubits' quantum state in the Bloch sphere spanned by $\{\ket{01},\ket{10}\}$. The eigenstates of $H_\text{2qb}$ are denoted by $\ket{\theta_+}$ and $\ket{\theta_-}$. At the end of (i) the qubit's state is $\ket{\psi(t_0)}$. After an unselective  measurement, the state is $\rho(t_0)$.  (c) Evolution of $\langle H_\text{2qb} \rangle$ (dotted brown), $\langle H_\text{loc} \rangle$ (dashed blue), and $\langle V \rangle$ (solid magenta) as a function of time (See text).
	}
	\label{fig1}
\end{figure}

Another core concept, quantum entanglement \cite{Erwin35}, was identified by Schr\"odinger as {\it the} characteristic trait of quantum physics.  This feature of quantum mechanics was originally identified by Einstein, Podolsky, and Rosen \cite{EPR} in their attempt to show quantum mechanics was incomplete and later derided by Einstein as ``spooky action at a distance". It has come to be viewed as an essential resource in various quantum technologies.  The spooky action is the consequence of wavefunction collapse, which happens because the measured non-local state is not an eigenstate of the local measured observable. In this Letter, we propose to exploit this feature to design a new generation of quantum measurement powered engines. Local measurements are performed on still interacting entangled systems, allowing to harvest the interaction energy. This contrasts with former entanglement engines powered with thermal resources \cite{Geraldine,Josefsson,Huang, Hewgill}. The measurement-based fueling mechanism we shall focus on also departs from the original EPR proposal, where the systems sharing the entangled state are space-like separated, and not interacting when the local measurements take place.

We first propose a bipartite engine made of two detuned qubits that become entangled through the coherent exchange of a quantum of excitation.  When the red-detuned qubit $A$ is initially excited, the excitation is partially transferred to the blue-detuned qubit $B$. Local energy measurements can then project the excitation into $B$ with a finite probability, resulting in some net energy gain \cite{jordan2004entanglement}. We provide evidence that this energy comes from the measurement channel, and corresponds to the cost of erasing the quantum correlations between the qubits. By exploiting the information carried by the measurement, one may extract this energy as work, in a cycle similar to the classical Szilard engine \cite{szilard1929entropieverminderung} or its quantum generalization \cite{Kim2011}. We demonstrate that in the limit of small detunings and large values of the couplings between the qubits, work extraction is nearly deterministic. Based on this mechanism of entanglement followed by a local measurement, we propose a protocol for frequency up-conversion over an $N$-qubit chain. Finally, we investigate the dynamics of the pre-measurement step, where one of the qubits is coupled with a quantum meter. Our analysis reveals a transfer of energy from the qubit-qubit correlations into the qubits-meter correlations, providing new insights into the physics of measurement-based engine fueling. \\


\textit{An entangled-qubits engine}---The basic mechanism of our engine is illustrated in Figs.~\ref{fig1}(a) and ~\ref{fig1}(b). It involves two qubits $A$ and $B$ of respective transition frequencies $\omega_A $ and $\omega_B$, whose evolution is ruled by the Hamiltonian
\begin{equation} \label{eq:H}
H_\text{2qb} = \sum_{i= A,B}\hbar\omega_i\sigma_i^\dagger\sigma_i +\hbar \frac{g(t)}{2}(\sigma_A^\dagger\sigma_B+\sigma_B^\dagger\sigma_A).
\end{equation}
We have introduced the lowering operator $\sigma_i = \ket{0_i}\bra{1_i}$ for the qubit $i \in\{A,B\}$.  
The first term of $H_\text{2qb}$ is the free Hamiltonian of the qubits. It thus features ``local" one-body terms that we shall denote as $H_\text{loc}$. The second term, which we denote by $V$, couples the qubits, giving rise to entangled states. The coupling channel can be switched on and off, which is modeled by the time-dependent coupling strength $g(t)$. In the rest of the paper we consider a positive detuning $\delta=\omega_B- \omega_A$. For simplicity, we denote the product states $\ket{x_A}\otimes\ket{y_B}$ as $\ket{xy}$, where $x,y \in \{0,1\}$.

The engine cycle encompasses four steps: (i) {\it Entangling evolution}. At time $t=0$, the qubits are prepared in the state $\ket{\psi_0} = \ket{10}$ of mean energy $\langle H_\text{2qb} \rangle = \bra{\psi_0}H_\text{2qb}\ket{\psi_0}= \hbar \omega_A$. The coupling term is switched on with a strength $g$. Since $\ket{\psi_0}$ is a product state, its mean energy does not change during this switching process, which is thus performed at no cost. The qubits' state then evolves into an entangled state $\ket{\psi(t)}$ where the initial excitation gets periodically exchanged between the two qubits, with 
\begin{align}
\ket{\psi(t)}=&(c_{\theta}^2 e^{i\Omega t/2}+s_{\theta}^2e^{-i\Omega t/2} )\ket{10} \nonumber  \\ 
&- c_{\theta}s_{\theta}( e^{i\Omega t/2} - e^{-i\Omega t/2} )\ket{01}.
\end{align}
We have defined $c_{\theta} = \cos(\theta/2)$, $s_{\theta} = \sin(\theta/2)$, $\theta $ as $\tan(\theta)=g/\delta$, and $\Omega=\sqrt{g^2+\delta^2 }$ the generalized Rabi frequency that characterizes the periodic energy exchange.

$\langle H_\text{loc} \rangle(t)$ and $\langle V \rangle(t)$ are plotted on Fig.~\ref{fig1}(c). As expected from a unitary evolution, their sum remains constant and equal to its initial value $\hbar \omega_A$. The periodic exchange of the single excitation between $A$ and $B$ gives rise to oscillations of the local energy component. This evolution is compensated by the opposite oscillations of the coupling energy $\langle V \rangle(t) \leq 0$. This term appears here as a binding energy of purely quantum origin, whose presence ensures that the total energy and the number of excitations are both conserved. 

 (ii) {\it Measurement}.  $\langle H_\text{loc}\rangle$ and $|\langle V \rangle(t)|$ reach a maximum when $t_0=\pi/\Omega$ where $\ket{\psi(t_0)} = i [\cos(\theta) \ket{10} - \sin(\theta) \ket{01}]$. At this time, a local projective energy measurement is performed on qubit $B$, and its outcome is encoded in a classical memory $M$. Here we consider an instantaneous process, performed with a classical measuring device. A more elaborate model of the measurement will be presented in the last part of the paper. In turn, the average qubits' state becomes a statistical mixture $\rho(\theta) =  \cos^2(\theta) \ket{10}\bra{10} + \sin^2(\theta)\ket{01}\bra{01}$,  erasing the quantum correlations between them and thus bringing the binding energy $\langle V(t_0)\rangle$ to zero.  The average energy input by the measurement channel is
 \begin{equation}\label{eq:W}
 E^\text{meas} =-\langle V(t_0)\rangle =\Delta \langle H_\text{2qb}\rangle= \hbar \delta \sin^2(\theta) \geq 0, 
 \end{equation}
where $\Delta \langle \cdot \rangle$ features the change of mean energy. Conversely, the von Neumann entropy of the qubit pair increases by an amount $S^\text{meas} = -\text{Tr}[\rho(\theta) \log_2(\rho(\theta))]$, that reads
 \begin{equation}\label{eq:S}
S^\text{meas} = -\cos^2(\theta) \log_2[\cos^2(\theta)] - \sin^2(\theta) \log_2[\sin^2(\theta)].
 \end{equation}
We use $\log_2$, such that all entropies are expressed in bits. The ratio ${\cal T^\text{meas}} =  E^\text{meas} /S^\text{meas}$ characterizes the measurement process from a thermodynamic standpoint. For a fixed detuning $\delta$, it diverges for large coupling where $\theta \rightarrow \pi/2$. It typically scales like ${\cal T^\text{meas}}\sim -\hbar \delta /[2(\pi/2-\theta)^2 \log_2 (\pi/2-\theta)]$. In this limit of large coupling and small detuning, quantum measurement can input a finite amount of energy with vanishing entropy. This contrasts with isothermal processes, where energy and entropy inputs are related by the bath temperature.

From an informational standpoint, the measurement creates classical correlations between the memory and the qubits states in the basis $\ket{10}, \ket{01}$. If the measurement is ideal, these correlations are perfect, such that the entropies of the qubits and the memory at the end of the process are equal. They are also equal to the mutual information they share, further denoted $I^\text{meas}(S:M)$. 

(iii) {\it Feedback}. The information stored in the memory is now processed to extract the energy input by the measurement. To do so, the coupling term is switched off at time ${t_0}^{+}$. Since the correlations between the qubits have been erased by the measurement, the switching-off can be implemented at no energetic cost. If the excitation is measured in $B$, which happens with probability $P_\text{succ} (\theta) = \sin^2(\theta)$, both $A$ and $B$ undergo a resonant $\pi$ pulse, such that $B$ emits a photon while $A$ absorbs one. The work $W=\hbar \delta$ is extracted and the qubits are reset to their initial state $\ket{10}$. Conversely if the excitation is measured in $A$, no pulse is implemented and the cycle restarts. Eventually, the mean work extracted is $W = E^\text{meas}$. At the end of this feedback step, the qubits' entropy vanishes, and a maximal amount of mutual information $|\Delta I(S:M)| = I^\text{meas}(S:M)$ is consumed.

(iv) {\it Erasure.} Immediately after the feedback, the memory's entropy still equals $S^\text{meas} = I^\text{meas}(S:M)$. The memory is finally erased in a cold bath, the minimal work cost of this operation being proportional to $S^\text{meas}$ \cite{LANDAUER}.

Since the whole cycle conserves the number of excitations, the states $\ket{01}$ and $\ket{10}$ of the two qubits feature an effective two-level system. This property allows us to picture the qubits dynamics in the Bloch sphere representation (Fig.~\ref{fig1}(b)) where the cyclic nature of the evolution becomes evident. 

The quantum engine described above extends the concept of measurement-fueled engines, originally proposed for single parties as working substances \cite{Elouard_maxwell, Elouard_Jordan,JEA,ding2018measurement,Yi2017}, to entangled systems. In those proposals the engine is fueled by quantum measurement back-action, which can only take place when the measured system state bears coherences in the basis of the measured observable. Both quantum measurement and coherence thus contribute to the fueling process. Similarly, in the present bipartite engine, both local measurements and entanglement are necessary for work extraction. \\

\begin{figure}[t]
	\centering
	\includegraphics[width=1\linewidth]{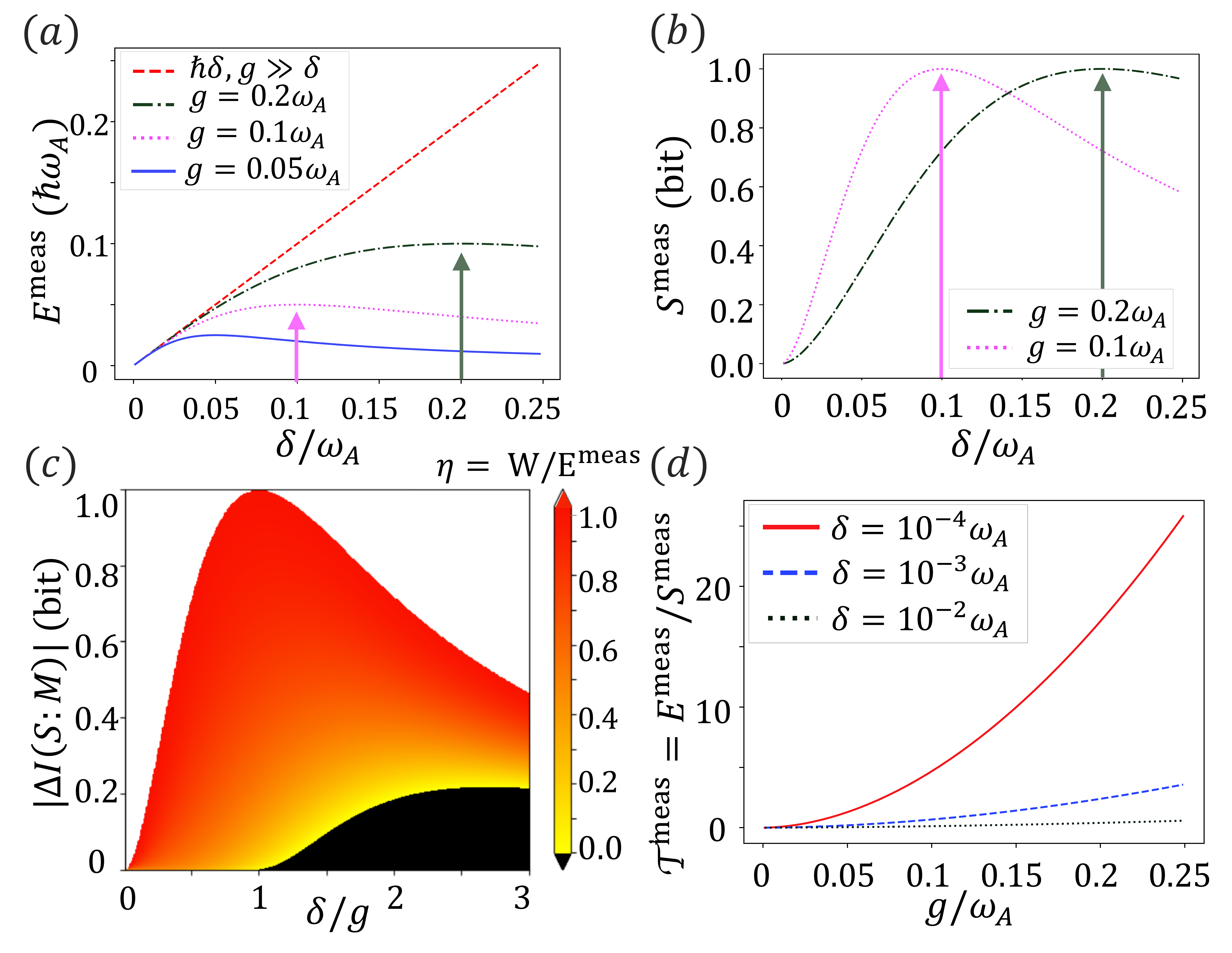}
	\caption{Measurement energy vs information as fuel. (a) Energy $E^{\text{meas}}$ and (b) entropy $S^{\text{meas}}$ inputs as a function of the detuning $\delta$, for various coupling strengths $g$. (c) Work extraction ratio $\eta = W/E^{\text{meas}}$ (color scale) as a function of $\delta/g$ and consumed mutual information $|\Delta I (S:M)|$. The black region corresponds to $\eta=0$.  d) Yield of information to work conversion ${\cal T}^{\text{meas}}$ as a function of $g$ for various $\delta$. 
	}
	\label{fig2}
\end{figure}

\textit{Measurement energy vs information as fuel}---The engine proposed above exploits two complementary features of quantum measurements: on the one hand, they bring energy and entropy, on the other hand, they extract information that can be further used to convert the energy input into work. Now focusing on the measurement and feedback steps, we analyze these energetic and informational resources, and how they respectively impact the performance of the bipartite engine.  

The mean energy $E^\text{meas}$ and entropy $S^\text{meas}$ input by the measurement process are plotted in Figs.~\ref{fig2}(a) and \ref{fig2}(b) as a function of the detuning $\delta$, for various coupling strengths $g$. As indicated in the figure, they are both maximized for $\delta = g$. This also corresponds to a maximal occupation of the memory and mutual information after the measurement step. 

Converting the measurement energy into work requires the processing of this information during the feedback step. The conversion is optimal ($W=E^\text{meas}$) when all information is consumed, which corresponds to the ideal cycle presented above. Non-optimal work extraction results from an incomplete consumption of information, $| \Delta I(S:M) | < I^\text{meas}(S:M)$, yielding a conversion ratio $\eta = W/E^\text{meas}<1$. We have modeled such an imperfect feedback in the Suppl. \cite{Suppl}. Figure~\ref{fig2}(c) features $\eta$ as a function of $\delta/g$ and $\Delta I(S:M)$, clearly showing the work value of information---the larger the consumed information, the larger the conversion ratio. Interestingly, the figure reveals that work can be extracted even if $\Delta I(S:M)=0$. This is the case when $P_\text{succ}(\theta)>1/2$, which happens when $\delta/g <1$. Then the $\pi$-pulses can be blindly applied, still leading to a net work extraction $W = \hbar \delta [\sin
^2(\theta)-\cos^2(\theta)]$. This mechanism solely exploits the energy input by the measurement, but not the extracted information; it is at play, e.g. in single temperature engines \cite{Yi2017, ding2018measurement}. By contrast, information processing is necessary when $\delta\geq g$. Note that in all non-ideal cases where information is not fully consumed, an additional step must be included in the cycle, to reset the qubits' state. 

From now on we suppose the feedback to be perfect, such that the information available in the memory is fully consumed and all the energy input by the measurement channel is converted into work.  In this situation, the net work extracted is $W=E^\text{meas}$. It is thus related to the size of the memory used $S^\text{meas}$ by the effective parameter ${\cal T^\text{meas}}$ defined above. Interestingly, now ${\cal T^\text{meas}}$ is a measure of efficiency of information-to-work conversion. Such efficiency is usually bounded by the bath temperature in Maxwell's demons, that are fueled by a thermal bath \cite{Parrondo2015, Masuyama2018}. ${\cal T^\text{meas}}$ is plotted on Fig.~\ref{fig2}(d) as a function of $g$ for various values of the detuning $\delta$. As it appears on the figure, it is not bounded and increases as a function of $g$. This reveals that in the limit $g \gg \delta$, a finite amount of work can be extracted by processing a vanishingly small amount of information. This effect is similar to the Zeno regime identified in Ref.~\cite{Elouard_maxwell}, where work extraction relies on measurements whose outcomes are nearly deterministic. \\

\begin{figure}[th!]
	\centering
	\includegraphics[width=1\linewidth]{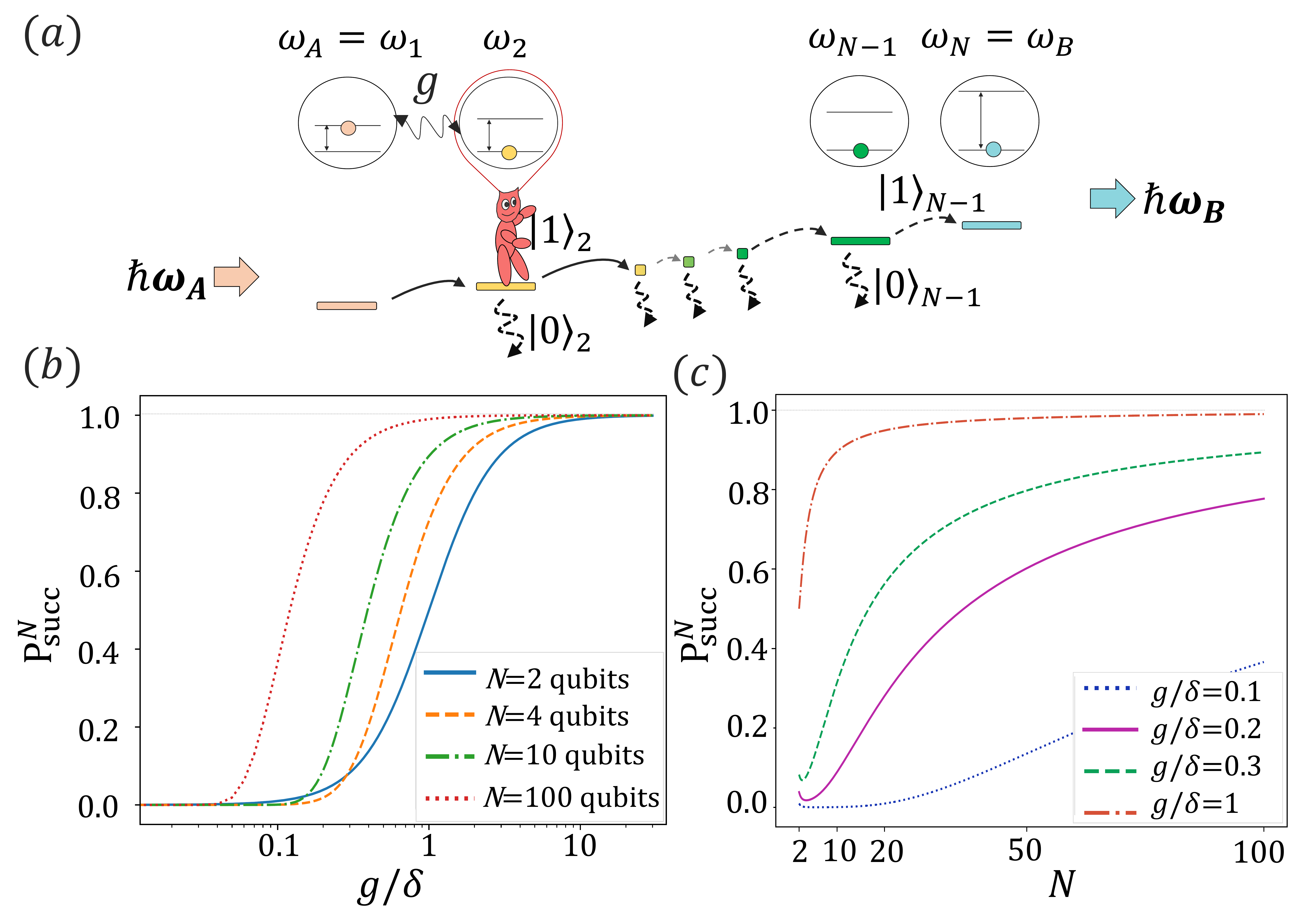}
	\caption{Entanglement and measurement based up-conversion mechanism. (a) Scheme of the frequency up-converter (See text). Probability of transfer $P_\text{succ}^N$ as a function of $g/\delta$ for various $N$ (b) and as a function of $N$ for various $g/\delta$ (c). The grey lines indicate constant values as guides to the eye.}
	\label{fig3}
\end{figure}

\textit{Up-conversion}--We now propose to exploit this mechanism to implement energy up-conversion.  The protocol is based on the efficient transfer of a single excitation through a chain of $N$ qubits of increasing frequency as depicted in Fig.~\ref{fig3}(a). We denote the frequency of the qubit $i$ by $\omega_i = \omega_A + (i-1)\delta/(N-1)$, with $i \in \{1,2,...,N\} $. As above, $\delta = \omega_B - \omega_A$, such that the frequency of qubit $N$ is $\omega_B$ and $\omega_1=\omega_A$. At time $t=0$, the qubit 1 
is excited and the coupling $g$ between qubit 1 and qubit 2 is switched on, its Rabi frequency being $\Omega_N = \sqrt{g^2+(\delta/(N-1))^2}$. At time $t_N = \pi/\Omega_N$,  the energy of qubit 2 is measured. The process stops if it is found in the ground state, which happens with probability $\cos^2(\theta_N)$, where $\tan(\theta_N) = (N-1)g/\delta= (N-1)\tan(\theta)$. If the excitation is successfully transferred to qubit 2, the coupling between $1$ and $2$ is switched off and the coupling between $2$ and $3$ is switched on. The same process is repeated between qubits $k$ and $k+1$ until the excitation gets detected in qubit $N$, which happens with probability $P_\text{succ}^{N} = \sin^{2(N-1)}(\theta_N)$. $P_\text{succ}^{N}$ is plotted in Fig.~\ref{fig3}(b) and \ref{fig3}(c) as a function of $g/\delta$ and $N$. For fixed values of $g$ and $\delta$, it is clearly advantageous to increase the number of intermediate qubits. The mechanism at play is reminiscent of the quantum Zeno effect. An analytic demonstration is presented in the Suppl. \cite{Suppl}. \\

\textit{Origin of the measurement fuel}---We finally investigate the measurement-based fueling mechanism, based on the modeling of the ``pre-measurement process" by which the qubits are entangled with a quantum meter while still coupled. It is well-known that such entanglement accounts for the entropy increase of the measured system. Below we show that it also explains the measurement energy input. 

\begin{figure}[th!]
	\centering
	\includegraphics[width=1\linewidth]{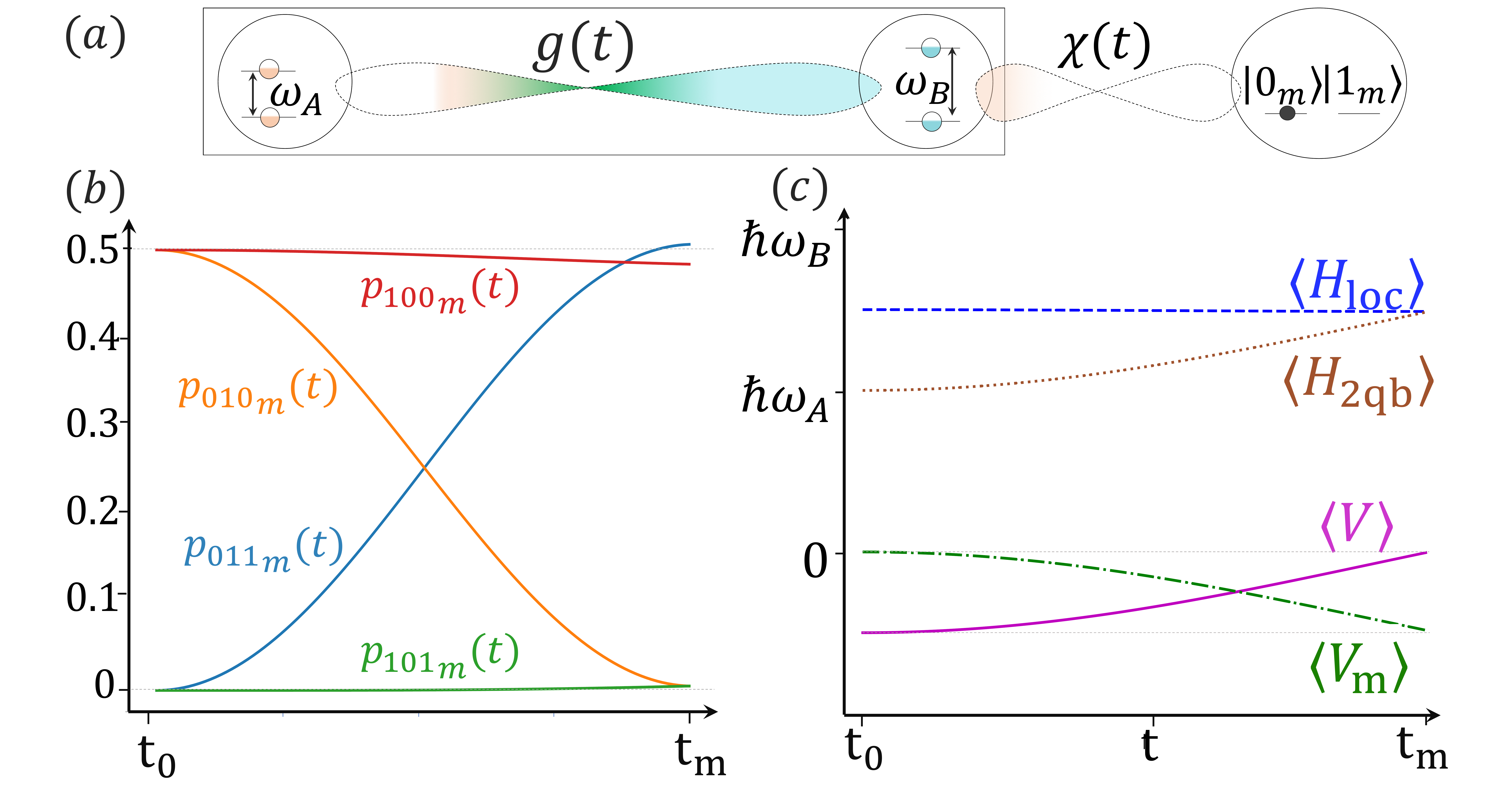}
	\caption{Dynamics of measurement-induced energy transfer. (a) Local quantum measurement of qubit $B$ allows for the creation of correlations between the meter $m$ and the $AB$ system and destroys correlations between the qubits. (b) Full state decomposition in the $\{\ket{100_m}, \ket{101_m}, \ket{010_m},\ket{011_m}\}$ basis during the pre-measurement step. (c)  Expectation values of $\langle H_\text{2qb} \rangle$, $\langle H_\text{loc} \rangle$, $\langle V_\text{m} \rangle$, and $\langle V \rangle$ as a function of the pre-measurement time $t \in [t_0, t_\text{m}] $.  The curves in the figure are calculated for $\chi = 10\Omega$ and $g=\delta$. The grey lines indicate constant values as guides to the eye. } 
	\label{fig4}
\end{figure}

The measurement process takes place between $t=t_0$ and $t_\text{m}$, and is depicted in Fig.~\ref{fig4}(a). The meter is chosen to be a third qubit $m$ with degenerate energy levels $\ket{0_m}$ and $\ket{1_m}$. It is coupled to the qubit $B$ through the Hamiltonian:

\begin{equation}
V_\text{m} = \hbar \chi(t) \sigma_B^\dagger\sigma_B\otimes \sigma_x^m.
\end{equation}
$\chi(t)$ is the measurement strength, with $\chi(t) = \chi$ for $t=[t_0, t_\text{m}]$ and $0$ otherwise. We choose $\chi \gg g$, to ensure the readout takes place on small time-scales with respect to the Rabi period. This defines the parameter $\epsilon = g/\chi$, which is small but finite since the measurement is implemented on still-interacting qubits.  

At $t_0^{-}$, the meter $m$ is prepared in $\ket{0_m}$, while $A$ and $B$ are in the entangled state $\ket{\psi(t_0)}$, such that their joint state reads $\ket{\Psi(t_0)} = i(\cos(\theta) \ket{100_m} - \sin(\theta) \ket{010_m})$. Since $\langle V_\text{m} (t_0)\rangle = 0$, the measurement channel is switched on at no energy cost. The joint qubits-meter system then evolves under the total Hamiltonian $H = H^{(0)} + H^{(1)}$, where $H^{(0)}= H_\text{loc}+V_\text{m}$ (resp. $H^{(1)}= V$) rules the evolution at zeroth order (resp. at first order) in the small parameter $\epsilon$. The evolution equations are solved at first order in the Suppl. \cite{Suppl}, yielding $\ket{\Psi^{(1)}(t)} = \ket{\Psi^{(0)}(t)}+\ket{\delta \Psi(t)}$ where $\ket{\delta \Psi(t)}$ is of order $\epsilon$.  The populations up to first order are plotted on Fig.~\ref{fig4}(b). To lowest order in $\epsilon$, the measurement is quantum non-demolition, resulting in state $ \ket{\Psi^{(0)}(t)} $ \cite{Grangier1998, Guryanova2020}. The readout is complete at time $t_\text{m}=t_0+ \pi/\chi$ where $ \ket{\Psi^{(0)}(t_\text{m})} = i[\cos(\theta) \ket{100_m} - \sin(\theta) \ket{011_m}]$. Conversely, the first order correction $\ket{\delta \Psi(t)}$ accounts for the remaining coupling between the qubits during the measurement. 

We now focus on energy flow and study the evolution of $\langle H_\text{loc} \rangle$, $\langle V\rangle$ and $\langle V_\text{m}\rangle$, see Fig.~\ref{fig4}(c). Since the process is unitary, these three components sum up to $\hbar \omega_A$. Perturbative calculations show that $\langle H_\text{loc} \rangle$ (resp. $\langle V\rangle$ and $\langle V_\text{m}\rangle$) remain constant up to first order in $\epsilon$ (resp. at zeroth order) \cite{Suppl}. The first order contribution of the binding energy between $A$ and $B$ reads $\langle V^{(1)} \rangle = \bra{\Psi^{(0)}(t)} V \ket{\Psi^{(0)}(t)}$, and thus scales like the coherences of the $AB$ density matrix in the $\ket{01},\ket{10}$ basis. Its absolute value decreases together with the quantum correlations between $A$ and $B$, and vanishes when the readout is complete. This evolution is compensated by an equivalent decrease of $\langle V_\text{m}^{(1)} \rangle (t)$, yielding at time $t_\text{m}$: $\langle V^{(1)}_\text{m}(t_\text{m}) \rangle = -E^{\text{meas}}$. Importantly, since $V_\text{m}$ scales as $\chi$, $\langle V^{(1)}_\text{m}(t_\text{m}) \rangle$ remains finite and of order $g$ even if $g/\chi \ll 1$. This calculation reveals the direction of the energy flow during the measurement process: The binding energy initially localized between the qubits is transferred between the qubits and the meter. This energy flow follows the same dynamics as the decoherence in the local energy basis, and can be seen as its energetic counterpart. Finally, when the readout is complete, the qubits-meter coupling must be switched off before any further operation can be done on the qubits. This switching off increases the qubits-meter energy by an amount $\langle -V_\text{m}(t_\text{m})\rangle = E^{\text{meas}}$. In the present non-autonomous scheme, this corresponds to the work cost paid to operate the measurement channel. \\

\textit{Outlook}---Our findings advance quantum measurement engines to encompass quantum entanglement and energy correlations, showing how entanglement engines may be powered by quantum measurement. From a conceptual standpoint, they shed new light on the measurement-based fueling process, and provide a unified view on former analyses based on analogies with work and heat exchanges. It should be recalled however that the concepts of work and heat were historically defined with respect to thermal noise and resources. Our results, on the other hand, are solely based on a stochasticity of quantum nature \cite{auffeves-grangier18}. They contribute to the emergence of a new framework---``Quantum energetics"---where thermodynamic concepts will be relevant in the presence of any kind of noise, especially at zero temperature where most quantum technology tasks are envisioned \cite{Landi-landauer-zero-T}.

In the future, it will be interesting to study the autonomous regimes of our engine where measurement and dissipation become time-independent processes, leading to the design of engines exploiting decoherence as a resource. This would bridge the gap with the field of dissipation engineering \cite{poya96,Kapit2017}, where dissipation is harnessed to produce nontrivial quantum states and desirable quantum dynamics. Such reservoir engineering has been recently employed in the circuit-QED architecture \cite{Liu16,Lu17,tou18,Ma2019}---the same experimental platform on which we expect to realize our proposed engine. \\

\begin{acknowledgments}
    We warmly thank M. Richard and C. Branciard for enlightening discussions. AA acknowledges the Agence Nationale de la Recherche under the Research Collaborative Project ``Qu-DICE" (ANR-PRC-CES47). P.A.C. acknowledges Templeton World Charity Foundation, Inc. which supported this work through the grant TWCF0338. K.M. acknowledges support from NSF  No. PHY-1752844 (CAREER) and the Research Corporation for Science Advancement.
\end{acknowledgments}

\end{document}